\documentclass[%
superscriptaddress,
preprint,
 amsmath,amssymb,
 aps,
prr,
longbibliography,
]{revtex4-1} %4-1

\usepackage{graphicx}% Include figure files
\usepackage{dcolumn}% Align table columns on decimal point
\usepackage{bm}% bold math
\usepackage{xcolor}
\usepackage{mathtools}
\usepackage{tabularx}
\usepackage[colorlinks=true,linkcolor=blue,allcolors=blue]{hyperref}

\begin{document}

%\preprint{APS/123-QED}

\title{Arbitrary-velocity laser pulses in plasma waveguides}

\author{J.P. Palastro}
\email{jpal@lle.rochester.edu}
\affiliation{
University of Rochester, Laboratory for Laser Energetics, Rochester, New York 14623-1299 USA}

\author{K.G. Miller}
\email{kmill@lle.rochester.edu}
\affiliation{
University of Rochester, Laboratory for Laser Energetics, Rochester, New York 14623-1299 USA}

\author{M. R. Edwards}
\affiliation{Department of Mechanical Engineering, Stanford University, Stanford, California 94305, USA}

\author{A.L. Elliott}
\affiliation{
University of Rochester, Laboratory for Laser Energetics, Rochester, New York 14623-1299 USA}

\author{L.S. Mack}
\affiliation{
University of Rochester, Laboratory for Laser Energetics, Rochester, New York 14623-1299 USA}

\author{D. Singh}
\affiliation{Department of Mechanical Engineering, Stanford University, Stanford, California 94305, USA}

\author{A.G.R. Thomas}
\affiliation{
G\'{e}rard Mourou Center for Ultrafast Optical Science, University of Michigan, Ann Arbor, Michigan 48109, USA}

\date{\today}

\begin{abstract}
Space-time structured laser pulses feature an intensity peak that can travel at an arbitrary velocity while maintaining a near-constant profile. These pulses can propagate in uniform media, where their frequencies are correlated with continuous transverse wavevectors, or in structured media, such as a waveguide, where their frequencies are correlated with discrete mode numbers. Here, we demonstrate the formation and propagation of arbitrary-velocity laser pulses in a plasma waveguide where the intensity can be orders of magnitude higher than in a solid-state waveguide. The flexibility to control the velocity of the peak intensity in a plasma waveguide enables new configurations for plasma-based sources of radiation and energetic particles, including THz generation, laser wakefield acceleration, and direct laser acceleration. 
\end{abstract}
     
\maketitle

\section{Introduction}
Space--time structuring of laser pulses has emerged as a powerful approach to tailoring light--matter interactions. The peak intensity of a space--time structured pulse can move independently of the group velocity—along, against, or transverse to the propagation direction—at any velocity, including those exceeding the vacuum speed of light \cite{Kondakci2017,Sainte-Marie2017,Froula2018,Kondakci2019,Li2020,Li2020a,Caizergues2020,palastro2020,jolly2020b,Yessenov2022,Liang23,Pigeon2024,Liberman24,Li2024,Gong2024,Guo2021,Huang2024,Su2025}. This flexibility has enabled novel configurations of light--matter interactions that have the potential to transform plasma-based applications \cite{Turnbull2018b,howard2019,Turnbull2019,Caizergues2020,palastro2020,Palastro2021, Ramsey2020,franke2021,Ramsey2022,Wu2022,Ye2023,Miller2023,Gong2024}, nonlinear optical processes \cite{Palastro2018, Kabacinski2023,Simpson2024,Simpson2024b,Fu2025}, and measurements of strong-field quantum electrodynamical phenomena \cite{dipiazza2021,formanek2022,formanek2024}. In a plasma, for instance, a laser pulse with a controllable velocity intensity peak can be used to overcome limitations on the energy gained by electrons in a laser wakefield accelerator \cite{Caizergues2020,palastro2020,Palastro2021,Miller2023}, ensure near-uniform conditions in a Raman amplifier \cite{Turnbull2018b,Turnbull2019,Wu2022}, or control the emission angle of high-power THz radiation \cite{Simpson2024,Simpson2024b,Fu2025}.

Optical techniques for space--time structuring reshape the amplitude, phase, or polarization of a laser pulse by imparting correlations between their spatial and temporal degrees of freedom. The resulting pulses can propagate through a uniform medium \cite{Kondakci2017,Sainte-Marie2017,Froula2018,Kondakci2019,Li2020,Li2020a,Caizergues2020,palastro2020,jolly2020b,Yessenov2022,Liang23,Pigeon2024,Liberman24,Gong2024,Li2024}, where the correlations are continuous, or through a structured medium, like a waveguide \cite{Guo2021,Huang2024,Su2025}, where the correlations are discrete. To date, the proposed applications for space--time structured pulses have been limited to uniform media. While waveguides could offer a complementary approach, the high intensities required for many applications would destroy a solid-state structure. Plasma waveguides \cite{Durfee1993,Ehrlich1996,Ditmire1998,Clark2000,Geddes2005,Kumarappan2005,Layer2007,Shalloo2018,Miao2020}, on the other hand, can withstand orders of magnitude higher intensities and are already used to guide conventional pulses in several of the applications envisioned for space--time structured laser pulses \cite{Geddes2004,Antonsen2007,Ren2007,Palastro2008, Pai2008,Turnbull2012,Hooker2014,Miao2022,Picksley2024}. 

\begin{figure}
\includegraphics[width=0.5\textwidth]{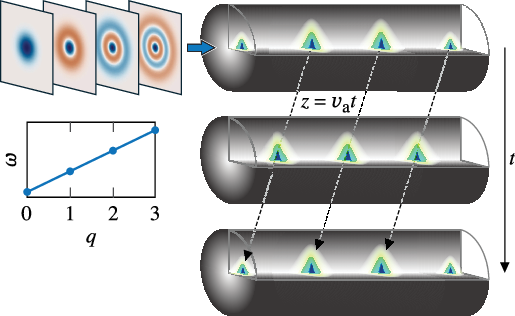}
\caption{A laser pulse composed of plasma-waveguide modes with appropriately selected frequencies exhibits an arbitrary-velocity intensity peak. In this example, the peak intensity travels backward with respect to the phase and group velocities of the modes. When the relative frequencies of the modes are commensurate (i.e., their ratios are rational), the moving intensity peak recurs at a regular interval.}
\label{fig:f1}
\end{figure}

Here, we demonstrate the propagation of high-intensity, arbitrary-velocity, space--time structured laser pulses in plasma waveguides. Figure \ref{fig:f1} illustrates the concept. A laser pulse composed of discrete waveguide modes with appropriately selected frequencies propagates through a preformed plasma channel. The interference of the modes produces an intensity peak that travels at a velocity that is independent of the modal phase and group velocities. When the ratios of the modal frequencies are rational, the intensity peak recurs at regular intervals, creating a train of arbitrary-velocity intensity peaks. This flexibility to control the velocity of the peak intensity in a plasma waveguide offers a new approach to high-intensity light--matter interactions that rely on velocity matching or extended interaction lengths. 

The remainder of this article begins with a model for the propagation of a laser pulse in a plasma waveguide  (Sec. II). The model is general enough to describe the guiding of a pulse with orbital angular momentum or vector vortex structure, regardless of the waveguide profile. The dispersion relation provided by the model is supplemented by a constraint that correlates the frequencies and wavenumbers of the modes composing the pulse (Sec. III). The constraint is chosen to produce a pulse with a peak intensity that moves independently of the group velocity at an arbitrary, specified value. The specific case of arbitrary-velocity laser pulses in a plasma waveguide with a parabolic density profile is analyzed (Sec. IV) and demonstrated with quasi-static particle-in-cell simulations (Sec. V). The article concludes with a summary of the results and a discussion of future prospects (Sec. VI). 

\section{Plasma Waveguides}

Consider a laser pulse propagating in the positive $\hat{\mathbf{z}}$ direction through a preformed plasma channel. The transverse electric field of the pulse can be expressed as a superposition of its frequency components:
\begin{equation}\label{eq:FT}
\mathbf{E}(\mathbf{x},t) = \frac{1}{4\pi}\int e^{-i\omega t}\tilde{\mathbf{E}}(\mathbf{x},\omega) d\omega +\mathrm{c.c.},
\end{equation}
where the integral is over positive frequencies. The frequency components evolve according to the wave equation
\begin{equation}\label{eq:waveeq}
\left(\frac{\partial^2}{\partial z^2} + \nabla^2_{\perp}+\frac{\omega^2}{c^2}\right)\tilde{\mathbf{E}}(\mathbf{x},\omega) = \frac{\omega_\mathrm{p}^2(\mathbf{x})}{c^2}\tilde{\mathbf{E}}(\mathbf{x},\omega),
\end{equation}
where $\omega_\mathrm{p}^2(\mathbf{x}) = e^2 n(\mathbf{x})/m_\mathrm{e}\varepsilon_0$ is the square of the plasma frequency and $n(\mathbf{x})$ is the electron density. The plasma channel is assumed to be longitudinally uniform, cylindrically symmetric, and underdense, such that $\omega_\mathrm{p}(\mathbf{x}) = \omega_\mathrm{p}(r) \ll \omega$, where $r=(x^2+y^2)^{1/2}$ is the transverse distance from the $z$ axis.

From the wave equation [Eq. \eqref{eq:waveeq}], one can derive the refractive index $\mu(r,\omega) = [1-\omega_\mathrm{p}^2(r)/\omega^2]^{1/2}$. This expression shows that an electron density that increases with radius results in a refractive index that decreases with radius. A refractive index that decreases with radius counteracts diffraction by bending the ``rays'' of the pulse towards the propagation axis ($r=0$). If the change in the refractive index is large enough, the plasma channel acts as a waveguide with bound states (i.e., eigenmodes) that remain transversely confined to the channel. 

The guided solutions to Eq. \eqref{eq:waveeq} are superpositions of these modes:
\begin{equation}\label{eq:sols}
\tilde{\mathbf{E}}(\mathbf{x},\omega) = \sum_{q,\ell} \tilde{\boldsymbol{\alpha}}_{q\ell}(\omega)A_{q\ell}(r)e^{ik_{q\ell}(\omega)z + i\ell \theta},
\end{equation}
where $q$ and $\ell$ are the radial and azimuthal mode numbers, respectively, $k_{q\ell}(\omega)$ is the wavenumber, and $\theta = \arctan{(y/x)}$ is the azimuth. The coefficients
\begin{equation}\label{eq:alpha}
\tilde{\boldsymbol{\alpha}}_{q\ell}(\omega) = |\tilde{\boldsymbol{\alpha}}_{q\ell}(\omega)|e^{i\Phi_{q\ell}(\omega)}\bold{\hat{e}}_{q\ell}(\omega)
\end{equation}
include the spectral amplitude $|\tilde{\boldsymbol{\alpha}}_{q\ell}(\omega)|$, spectral phase $\Phi_{q\ell}(\omega)$, and  polarization unit vector $\bold{\hat{e}}_{q\ell}(\omega)$ of each mode. The radial functions satisfy 
\begin{equation}\label{eq:operator}
\left[\nabla^2_{\perp}-\frac{\omega_\mathrm{p}^2(r)-\omega_\mathrm{p0}^2}{c^2}\right]A_{q\ell}(r) = -\frac{1}{w_{q\ell}^2}A_{q\ell}(r),
\end{equation}
where $\omega_\mathrm{p0}^2 \equiv \omega_\mathrm{p}^2(0)$. The $w^{-2}_{q\ell}$ take discrete, positive-definite scalar values, and the transverse width of each mode is $\propto w_{q\ell}$. The exact form of the $A_{q\ell}(r)$ and $w_{q\ell}$ depend on the profile of the channel $\omega_\mathrm{p}^2(r)$.  Combining Eqs. \eqref{eq:waveeq}, \eqref{eq:sols}, and \eqref{eq:operator} provides the modal dispersion relation
\begin{equation}\label{eq:dispersion}
c^2k^2_{q\ell} = \omega^2 -\omega_\mathrm{p0}^2 - \frac{c^2}{w_{q\ell}^2}.
\end{equation}
Thus, the $w_{q\ell}^{-1}$ contribute to the dispersion relation like a transverse wavevector.

The phase and group velocities of each mode can be calculated from the modal dispersion relation. The phase velocities $v_{\phi,q\ell} = \omega/k_{q\ell}$ are faster than the vacuum speed of light (i.e., superluminal) and given by
\begin{equation}\label{eq:vphase}
v_{\phi,q\ell} = c\left(1 - \frac{c^2}{\omega^2 w_{q\ell}^2} - \frac{\omega_\mathrm{p0}^2}{\omega^2}\right)^{-1/2}.
\end{equation}
The group velocities $v_{\mathrm{g},q\ell} = \partial \omega / \partial k_{q\ell}$ are slower than the vacuum speed of light (i.e., subluminal) and given by
\begin{equation}\label{eq:vgroup}
v_{\mathrm{g},q\ell} = c\left(1 - \frac{c^2}{\omega^2 w_{q\ell}^2} - \frac{\omega_\mathrm{p0}^2}{\omega^2}\right)^{1/2}.
\end{equation}
The subluminal group velocities can limit or preclude laser--plasma-based applications that rely on matching the velocity of the peak intensity to some underlying process.

\section{Arbitrary-Velocity Guiding}

By appropriately selecting the frequency of each mode, a guided laser pulse can exhibit an intensity peak that moves independently of the modal group velocities $v_{\mathrm{g},q\ell}$. While each mode must satisfy the dispersion relation  [Eq. \eqref{eq:dispersion}], an additional constraint on the frequencies and mode numbers can also be imposed. In the specific case of interest here, this constraint is chosen so that the superposition of modes travels at an arbitrary group velocity $v_\mathrm{a}$. 

To determine the constraint necessary for an arbitrary group velocity $v_\mathrm{a}$, one can directly integrate $\partial \omega / \partial k_{q\ell} = v_\mathrm{a}$ to find $\omega = v_\mathrm{a}k_{q\ell} + \eta$. The integration constant $\eta$ is set by requiring that $\omega = \omega_{00} \equiv (\omega_\mathrm{p0}^2 + c^2/w_{00}^2 + c^2k_{00}^2)^{1/2}$ when $k_{q\ell} = k_{00}$. The resulting constraint is 
\begin{equation}\label{eq:constraint}
\omega =  \omega_{00} + v_\mathrm{a}(k_{q\ell}-k_{00}).
\end{equation}
Note that the integration constant can be set so that $\omega = \omega_{pl}$ when $k_{q\ell} = k_{pl}$ for any $p$ and $l$ without affecting the arbitrary group velocity. The lowest-order mode $p = l = 0$ is used here because it provides a convenient reference frequency. For notational brevity, $\omega_{00}$, $k_{00}$, $v_{\phi,00}$, and $v_{\mathrm{g},00}$ will be shortened to $\omega_{0}$, $k_{0}$, $v_{\phi0}$, and $v_{\mathrm{g}0}$  hereafter.

\begin{figure}
\includegraphics[width=0.5\textwidth]{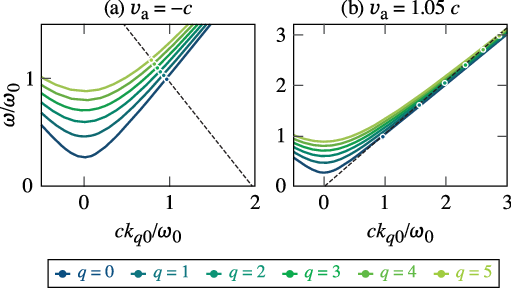}
\caption{To produce an arbitrary-velocity intensity peak in a plasma waveguide, the frequency of each mode composing the laser pulse must lie at the intersection of the modal dispersion relation [Eq. \eqref{eq:dispersion}] and the constraint $\omega = \omega_0 + v_\mathrm{a}(k_{q\ell}-k_0)$.  (a) The intersections for $v_\mathrm{a} = -c$ and $\ell = 0$. (b) The intersections for $v_\mathrm{a} = 1.05c$ and $\ell = 0$. An artificially small matched spot size $\omega_0w/c = 7.5$ was used for this figure to make the intersection points more visible. The other parameters are provided in Table I and are identical to those used in the simulations.}
\label{fig:f2}
\end{figure}

The frequency and wavenumber pairs that satisfy both the dispersion relation and the constraint are determined by the intersections of Eqs. \eqref{eq:dispersion} and \eqref{eq:constraint} (see Fig. \ref{fig:f2}). The frequencies at which the curves intersect, denoted by $\omega_{q\ell}$, are found by inserting Eq. \eqref{eq:constraint} into Eq. \eqref{eq:dispersion}. This yields a quadratic equation in $\omega$ with the solutions 
\begin{equation}\label{eq:omgql}
\frac{\omega_{q\ell}}{\omega_{0}} = 1 - \frac{v_\mathrm{a}(v_{\mathrm{g}0}-v_\mathrm{a})}{c^2-v_\mathrm{a}^2} \pm \left[\frac{v_\mathrm{a}^2(v_{\mathrm{g}0}-v_\mathrm{a})^2}{(c^2-v_\mathrm{a}^2)^2} - \frac{v_\mathrm{a}^2W_{q\ell}^{-2}}{(c^2-v_\mathrm{a}^2)} \right]^{1/2},
\end{equation}
where $W_{q\ell}^{-2} \equiv (c/\omega_0)^2(w_{q\ell}^{-2} - w_{00}^{-2})$. The $\pm$ corresponds to the two possible intersections of a line [Eq. \eqref{eq:constraint}] and a hyperbola [Eq. \eqref{eq:dispersion}]. Figure \ref{fig:f2} shows an example of the intersections for $v_\mathrm{a} = -c$ and $v_\mathrm{a} = 1.05c$. In both cases, the minus sign in Eq. \eqref{eq:omgql} corresponds to intersection points with positive frequency and wavenumber.

The constraint that establishes the arbitrary group velocity can be built into the electric field $\mathbf{E}$ through the frequency dependence of the $\tilde{\boldsymbol{\alpha}}_{q\ell}(\omega)$. More specifically, the $\tilde{\boldsymbol{\alpha}}_{q\ell}(\omega)$ should be sharply peaked around $\omega = \omega_{q\ell}$. In the simplest case, the spectral amplitudes for each mode have the same sharply peaked profile $|\tilde{\boldsymbol{\alpha}}_{q\ell}(\omega)| = \tilde{\alpha}(\omega-\omega_{q\ell})$; the spectral phases are independent of frequency $\partial_\omega \Phi_{q\ell} = 0$; and all modes share a common, frequency-independent polarization $\bold{\hat{e}}_{q\ell}(\omega) = \hat{\mathbf{x}}$. Upon applying these simplifications, substituting Eq.~\eqref{eq:sols} into Eq.~\eqref{eq:FT}, and Taylor expanding $k_{q\ell}(\omega)$ about $\omega = \omega_{q\ell}$, one finds the transverse electric field of an arbitrary-velocity guided pulse:
\begin{equation}\label{eq:Ex}
\begin{aligned}
E_x(\mathbf{x},t) &\approx \frac{1}{2}e^{-i\omega_0(t - z/v_{\phi0}) }\sum_{q,\ell} \alpha(t-z/v_{\mathrm{g},q\ell})A_{q\ell}(r)  \\ 
&\exp\left[-i\Omega_{q\ell}(t-z/v_\mathrm{a}) + i\ell\theta +i\Phi_{q\ell} \right] +\mathrm{c.c.},
\end{aligned}
\end{equation}
where $\Omega_{q\ell}\equiv \omega_{q\ell} - \omega_0$ and $\alpha(t) = \tfrac{1}{2\pi}\int \tilde{\alpha}(\omega)e^{-i\omega t}d\omega$. 

The transverse electric field of the guided pulse features three velocities. The phase fronts of the overall field travel at the nominal phase velocity of the lowest-order mode $v_{\phi0}$. The envelope of each mode composing the field $\alpha$ travels at the group velocity of that mode $v_{\mathrm{g},q\ell}$. It is this velocity that determines the speed of energy transport and ensures causality is not violated when $|v_\mathrm{a}| > c$. Finally, and most importantly, the phase terms $\propto \Omega_{q\ell}(t-z/v_\mathrm{a})$ produce a time-dependent interference pattern with an intensity peak that travels at $v_\mathrm{a}$. The properties of this intensity peak depend on the specific density profile of the plasma channel.

\begin{figure}
\includegraphics[width=0.5\textwidth]{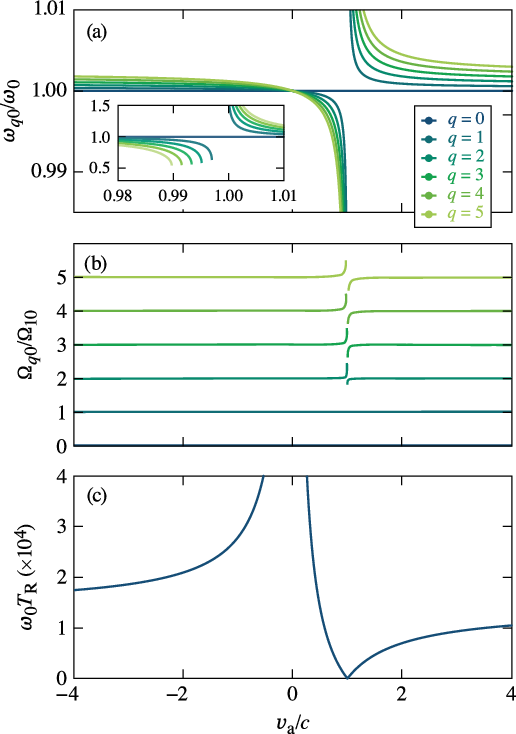}
\caption{(a) The frequencies of each mode needed to produce an intensity peak traveling at a velocity $v_\mathrm{a}$. The inset shows the behavior close to $v_\mathrm{a} = c$ and the small range of prohibited velocities [Eq. \eqref{eq:unobtain}]. (b) The relative frequency ratios needed to produce an intensity peak traveling at $v_\mathrm{a}$. When the relative frequency ratios are integer, the intensity peak recurs at a regular interval $T_\mathrm{R}$.  (c) The recurrence period $T_\mathrm{R}$ as a function of $v_\mathrm{a}$.}
\label{fig:f3}
\end{figure}

Before moving on to a specific profile, it is pertinent to examine general features of the intersection frequencies $\omega_{q\ell}$. First, when $v_\mathrm{a} = 0$, the intersection frequency of every mode is the same---that is $\omega_{q\ell} = \omega_0$ and $\Omega_{q\ell} = 0$ [Fig. \ref{fig:f3}(a)]. With $\Omega_{q\ell} = 0$, there is no interference between the modes and no discernible moving intensity peak other than that of the overall profiles $\alpha(t-z/v_{\mathrm{g},q\ell})$. The modes must have distinct frequencies in order to supply the bandwidth necessary for a finite-duration intensity peak.

Second, the radical in Eq. \eqref{eq:omgql} is negative for some values of $v_\mathrm{a}$, which indicates that each mode has a range of velocities that are prohibited. Upon setting the radical to zero, one finds that the ranges of prohibited velocities are given by
\begin{equation}\label{eq:unobtain}
\left|v_\mathrm{a} - \frac{v_{\mathrm{g}0}}{1+W_{q\ell}^{-2}}\right| < \frac{cW_{q\ell}^{-1}}{1+W_{q\ell}^{-2}} \left(1 - \frac{v_{\mathrm{g},q\ell}^2}{c^2} \right)^{1/2}. 
\end{equation}
For typical parameters, $W_{q\ell}^{-2}\ll1$ and $v_{\mathrm{g},q\ell} \approx 1$, such that the right-hand side of Eq. \eqref{eq:unobtain} is very small. Thus, the prohibited values of $v_\mathrm{a}$ are limited to a small subluminal interval about $v_{\mathrm{g}0}$ [see Fig. \ref{fig:f3}(a) inset]. 

Finally, when $W_{q\ell}^{-2} (v_\mathrm{a} - v_{\mathrm{g}0})^{-2}|c^2-v_\mathrm{a}^2| \ll 1$, both roots in Eq.~\eqref{eq:omgql} reduce to the simpler ``paraxial'' form
\begin{equation}\label{eq:parax}
\frac{\omega_{q\ell}}{\omega_{0}} \approx
1 + \frac{v_\mathrm{a}W_{q\ell}^{-2}}{2(v_\mathrm{a} - v_{\mathrm{g}0})}. 
\end{equation}
This is generally a good approximation except for a small range of $v_\mathrm{a}$ values about $v_{\mathrm{g}0}$, which is largely prohibited anyway [Eq. \eqref{eq:unobtain}]. In the special cases of $v_\mathrm{a} = \pm c$, the approximate equality in Eq. \eqref{eq:parax} becomes an exact equality. The paraxial form of $\omega_{q\ell}$ is useful for analyzing the interference pattern and mode beating of the guided pulse for specific electron density profiles. An explicit expression for Eq. \eqref{eq:parax} in terms of physical parameters for the specific case of a parabolic plasma channel can be found in Appendix A. 

\begin{table}[b]
\caption{\label{tab:table1}
Laser pulse and plasma parameters used for the figures and simulations. The parameters are motivated by commonly used laser systems and experimentally demonstrated plasma channels.  In the rightmost column, space, time, and density are normalized to $c/\omega_{0}$, $1/\omega_{0}$, and $n_\mathrm{cr} = m_\mathrm{e}\varepsilon_0\omega_0^2/e^2$. The vacuum wavelength $\lambda_0 = 2\pi c/\omega_0$.}
%\begin{ruledtabular}
\begin{tabular}{c c c}
\hline
\hline
Pulse parameters & Value & Normalized \\
\hline
$\lambda_0$ & 1 $\mu$m & 2$\pi$ \\
$\omega_0$ & $1.9\times10^{15}$ rad/s & 1 \\
$a_0$ & 0.6 & 0.6 \\
$T$  & 7.7 ps & 15000 \\
$q_\mathrm{max}$  & 5 & 5 \\
\hline
Plasma parameters & Value & Normalized \\
\hline
$n(0)$ & $1\times10^{18} \text{cm}^{-3}$ & 9.0$\times10^{-4}$ \\
$w$ & $15$ $\mu$m & 94 \\
\hline
\hline
\end{tabular}
%\end{ruledtabular}
\end{table}

\section{Parabolic Plasma Channel}

Two common techniques for creating plasma channels rely on the hydrodynamic expansion of a laser-heated plasma. The original technique employs long (ns) pulses where the heating is dominated by inverse-bremsstrahlung \cite{Durfee1993}. A more recent technique employs shorter pulses (ps) where the heating is dominated by the residual energy from above-threshold photoionization \cite{Shalloo2018,Miao2020}. In both cases, the hydrodynamic expansion produces a plasma with a locally parabolic transverse profile. 

The profile of a parabolic plasma channel can be parameterized as follows:
\begin{equation}\label{eq:parab}
\omega_\mathrm{p}^2(r) = \omega_{\mathrm{p}0}^2 + \frac{4c^2}{w^4}r^2, 
\end{equation}
where $w$ is the ``matched'' spot size. The corresponding $A_{q\ell}(r)$ and $w_{q\ell}^{-2}$ are given by
\begin{align}\label{eq:Apara}
A_{q\ell}(r) &= \bigg(\frac{\sqrt 2 r}{w} \bigg)^{|\ell|}L_{q}^{|\ell|}\bigg(\frac{2r^2}{w^2} \bigg)\mathrm{exp}\bigg(-\frac{r^2}{w^2}\bigg),  \\
\frac{1}{w_{q\ell}^2} &= \frac{4}{w^2}(1 + 2q + |\ell|),\label{eq:wpara}
\end{align}
where $L_q^{|\ell|}$ is a generalized Laguerre polynomial. This expression for $w_{q\ell}^{-2}$ yields $W_{q\ell}^{-2} = (2c/\omega_0w)^2(2q+|\ell|)$. While parabolic plasma channels allow for arbitrary-velocity intensity peaks with any or multiple values of $\ell$, the remainder of this work will focus on the case of $\ell = 0$. 

Figure \ref{fig:f3}(a) displays the intersection frequencies $\omega_{q0}$ for a parabolic channel as a function of $v_\mathrm{a}$ (see Table I for parameters). The frequencies are spaced relatively closely except for a narrow range of velocities about $v_\mathrm{a} = c$. The inset highlights the behavior in this range. To the left of $v_\mathrm{a} = c$, the $\omega_{q0}$ curves terminate at the lower bounds of the prohibited ranges [Eq. \eqref{eq:unobtain}]. To the right of $v_\mathrm{a} = c$, the curves continue rising as they approach the subluminal upper bounds of the prohibited ranges. The frequencies are identical at $v_\mathrm{a} = 0$.

A parabolic plasma channel has the noteworthy property that the relative frequencies $\Omega_{q\ell}$ are approximately harmonic.  This property results in recurrences of the moving intensity peak at regular intervals. To illustrate the harmonicity, Fig. 3(b) shows the ratios $\Omega_{q0}/\Omega_{10}$. Over the range of velocities where the paraxial approximation [Eq. \eqref{eq:parax}] is valid, $\Omega_{q0}/\Omega_{10} \approx q$, whereas when $v_\mathrm{a} = \pm c$, $\Omega_{q0}/\Omega_{10} = q$. Thus, the relative frequencies are nearly harmonic everywhere except for the small range of subluminal velocities about $v_\mathrm{g0}$.

The recurrences and other properties of the moving intensity peaks can be inferred from the on-axis intensity of the guided pulse $I$. In the paraxial approximation,
\begin{equation}\label{eq:Inten}
\begin{aligned}
I(z,t) &\approx I_0(t-z/v_\mathrm{g0}) \\ & \Bigg| 1+ \sum_{q=1}^{q_\mathrm{max}} \mathrm{exp}\left[-iq\Omega_{10}(t-z/v_\mathrm{a}) + i\Phi_{q\ell} \right] \Bigg|^2,
\end{aligned}
\end{equation}
where $\Omega_{10} = 4c^2v_\mathrm{a}/[(v_\mathrm{a}-v_\mathrm{g0})\omega_0w^2]$ and $\alpha \approx (2I_0/c\varepsilon_0)^{1/2}$ has been used. The summation in Eq. \eqref{eq:Inten} is a truncated Fourier series. As a result, the intensity peak formed by the interference of the waveguide modes recurs with a period $T_\mathrm{R} = 2\pi q/\Omega_{q0}$ or
\begin{equation}\label{eq:Trecur}
T_\mathrm{R} = \frac{\pi \omega_0 w^2}{2c^2} \bigg| \frac{v_\mathrm{a} - v_\mathrm{g0}}{v_\mathrm{a}} \bigg|
\end{equation}
[see Fig. \ref{fig:f3}(c)]. The corresponding recurrence distance is $L_\mathrm{R} = T_\mathrm{R}/v_\mathrm{a}$. The effective duration of the moving intensity peaks roughly scales as
\begin{equation}\label{eq:Tdura}
\tau \sim \frac{\pi \omega_0 w^2}{4c^2 q_\mathrm{max}} \bigg| \frac{v_\mathrm{a} - v_\mathrm{g0}}{v_\mathrm{a}} \bigg|.
\end{equation}
The velocity-dependence of the duration is consistent with the spread in frequencies (i.e., the bandwidth) observed in Fig. \ref{fig:f3}. With each additional mode, the bandwidth increases, leading to the scaling $\tau \propto 1/q_\mathrm{max}$. The maximum distance $L$ that an intensity peak can travel is determined by its transit time through the entire laser pulse. For an intensity profile $I_0$ (or $\alpha$) of duration $T$,
\begin{equation}\label{eq:L}
L =  \bigg| \frac{v_\mathrm{g0}v_\mathrm{a}}{v_\mathrm{g0} - v_\mathrm{a}} \bigg| T. 
\end{equation}
Together with Eq. \eqref{eq:Trecur}, this results in the velocity-independent relation $T_\mathrm{R}/T = \pi w^2 v_\mathrm{g0}/ (c^2 \omega_0 L)$. Note that a recurrence cannot occur if $T<T_R$, which can be used to isolate a single moving intensity peak. 

The prominence of recurrences and scaling of the effective duration $\tau \propto q^{-1}_\mathrm{max}$ depend on the specific profile of the plasma channel $\omega_\mathrm{p}^2(r)$. For instance, in a step-index channel, the relative frequencies required for an arbitrary-velocity intensity peak are not harmonic (see Appendix B). The relative frequencies are, however, nearly commensurate when $\ell = 0$, i.e., their ratios are approximate rational numbers. This results in a recurring intensity peak surrounded by less-structured, lower-amplitude modulations. In addition, quartic or higher-order radial perturbations to a parabolic plasma channel introduce anharmonic corrections to the relative frequencies. This causes the recurrences to phase mix away over sufficiently long propagation distances (see Appendix C).

\begin{figure}
\includegraphics[width=0.9\textwidth]{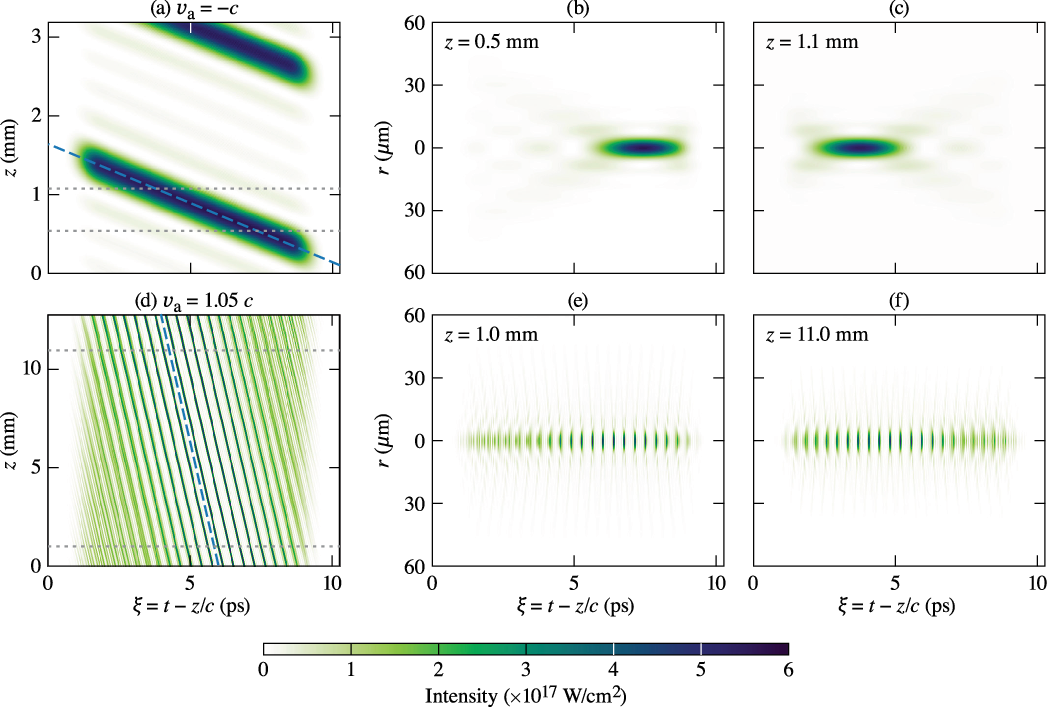}
\caption{Evolution of arbitrary-velocity pulses in a plasma waveguide for (a)--(c) $v_\mathrm{a} = -c$ and (d)--(f) $v_\mathrm{a} = 1.05c$. (a,d) The on-axis ($r=0$) intensity as a function of the moving-frame coordinate $\xi = t - z/c$ and propagation distance $z$. The dashed blue lines show the theoretical predictions $z = v_\mathrm{a} t = cv_\mathrm{a}\xi/(c-v_\mathrm{a})$. At a fixed $\xi$, the intensity recurrences appear after every ${\approx} 2.2$ mm of propagation distance. (b), (c), (e), and (f) Intensity profiles at the two propagation distances marked by the gray dashed lines in (a) and (d). The moving intensity peaks maintain their spatiotemporal profile as they traverse the focal range $L = |v_{\mathrm{g}0}v_\mathrm{a}/(v_{\mathrm{g}0}-v_\mathrm{a})|T$. Note that the use of $\xi = t - z/c$ as the abscissa gives the impression that the $v_\mathrm{a} = -c$ intensity peak moves forward despite its negative velocity.}
\label{fig:f4}
\end{figure}

\section{Simulation Results}

The analysis above describes the construction and features of arbitrary-velocity laser pulses in a plasma waveguide. This analysis, however, is limited to linear propagation. Many of the envisioned applications of arbitrary-velocity pulses require high intensities (${>} 10^{16} \; \mathrm{W/cm^2}$), where the response of the plasma and its effect on the pulse depend nonlinearly on the amplitude of the pulse. In fact, the degree of nonlinearity is quantified by the amplitude, or maximum normalized vector potential ${a_0 = 8.5\times10^{-10} \lambda_0 [\mu\mathrm{m}] (I_\mathrm{M} [\mathrm{W/cm^2}])^{1/2}}$, where $\lambda_0 = 2\pi c /\omega_0$ and $I_\mathrm{M}$ is the maximum intensity. To demonstrate the propagation of mildly nonlinear ($a_0 = 0.6$, $I_\mathrm{M} = 5\times10^{17} \; \mathrm{W/cm^2}$) arbitrary-velocity pulses in a plasma waveguide,  simulations were conducted using the quasi-static particle-in-cell code \textsc{qpad} (see Table I for physical parameters and Appendix D for numerical details).

Figure \ref{fig:f4} displays the simulation results for $v_\mathrm{a} = -c$ (top) and $v_\mathrm{a} = 1.05c$ (bottom). In both cases, the intensity peaks travel at the designed velocity $v_\mathrm{a}$, recur after the predicted times $T_\mathrm{R}$ and propagation distances $L_\mathrm{R}$, and have the expected durations $\tau$. The simulations use the moving-frame coordinate $\xi = t - z/c$ in place of $t$, such that the designed trajectories are given by 
\begin{equation}\label{eq:L}
z_\mathrm{a} = v_\mathrm{a} (t-t_\mathrm{i}) = \frac{cv_\mathrm{a}}{c-v_\mathrm{a}}(\xi-t_\mathrm{i}),
\end{equation}
where $t_\mathrm{i}$ is the temporal location of an intensity peak at $z_\mathrm{a}=0$ [blue dashed lines in Figs. \ref{fig:f4}(a) and (d)]. Note that the use of the moving-frame coordinate gives the impression that the $v_\mathrm{a} = -c$ intensity peak is moving forward despite its negative velocity ($\partial_\xi z_\mathrm{a} < 0$ for both $v_\mathrm{a} > 1$ and $v_\mathrm{a}<0)$.

Figures \ref{fig:f4}(a) and (d) illustrate the dependence of the recurrence time and distance on the velocity $v_\mathrm{a}$. For the simulated parameters, the $v_\mathrm{a} = -c$ recurrence time is $T_\mathrm{R} = 15$ ps, which is nearly twice the duration of the entire pulse $T = 7.7$ ps. As a result, only one intensity peak is visible in $\xi = t - z/c$ [Fig. \ref{fig:f4}(a)]. This contrasts the case of $v_\mathrm{a} = 1.05c$ where the recurrence time $T_\mathrm{R} = 380$ fs is much shorter than the pulse duration and many peaks are visible in $\xi$ [Fig. \ref{fig:f4}(d)]. In the moving-frame coordinate system, the recurrence distance is given by $L_\mathrm{R} = cT_\mathrm{R}|v_\mathrm{a}/(v_\mathrm{a} - v_\mathrm{g0})| = \pi\omega_0w^2/2c$. Thus, both the $v_\mathrm{a} = -c$ and $v_\mathrm{a} = 1.05c$ peaks recur after $L_\mathrm{R} = 2.2$ mm.

Figures \ref{fig:f4}(b), (c), (e), and (f) show the radial profiles of the $v_\mathrm{a} = -c$ and $v_\mathrm{a} = 1.05c$ pulses at the propagation distances indicated by the horizontal gray lines in (a) and (d). In agreement with Eq. \eqref{eq:Tdura}, the effective duration $\tau$ of the $v_\mathrm{a} = 1.05c$ peaks is $20\times$ shorter than the effective duration of the $v_\mathrm{a} = -c$ peak. Figures \ref{fig:f4}(e) and (f) also show the effect of the frequencies being slightly anharmonic due to deviations from the paraxial approximation [Eq. \eqref{eq:parax}]. The intensity peaks at the left and right edges of the pulse appear more diffuse because the spatial modes composing the pulse have slipped out of phase. Surprisingly, the principal features of the arbitrary-velocity intensity peaks appear resistant to nonlinear propagation. The intensity limit of this resistance will be a subject for future study.

\section{Conclusions and Prospects}
A high-intensity laser pulse propagating through a plasma waveguide can exhibit an intensity peak that travels at an arbitrary velocity. The pulse is constructed by superposing modes of the waveguide with appropriate frequencies. The peak intensity can travel subluminally, superluminally, or backward with respect to the phase and group velocities of the modes. When the modal frequencies are commensurate (i.e., their ratios are rational numbers), the intensity peak recurs at regular intervals. The construction, propagation, and properties of these arbitrary-velocity pulses were analyzed theoretically and demonstrated with quasi-static particle-in-cell simulations.

The simulations used laser pulse and plasma parameters motivated by commonly used laser systems and experimentally demonstrated plasma channels. A laser pulse composed of spatial modes with different frequencies can be assembled using 
recently developed techniques for spatiotemporal structuring of laser pulses \cite{Cruz-Delgado2022,Piccardo2023,Zhan24}. For instance, a broad bandwidth pulse could be dispersed into several ``pulselets" with different central frequencies. Each pulselet could then be spatially and spectrally structured using metasurface optics or a spatial light modulator \cite{Cruz-Delgado2022,Piccardo2023,Zhan24}. Alternatively, the multiplexed pulses of a high-power fiber laser \cite{Rainville2024} could be independently manipulated with these same optics before being coherently combined. With recent advances in plasma optics \cite{Edwards2022}, it may also be possible to split different frequency bands of a high-power laser pulse into distinct spatial modes using an appropriately designed diffractive plasma element.

The realization of arbitrary-velocity laser pulses in a plasma channel would offer new possibilities for laser--plasma-based applications, including laser wakefield acceleration \cite{Hooker2014,Caizergues2020,palastro2020}, THz generation \cite{Antonsen2007}, and direct laser acceleration \cite{Palastro2008}. In laser wakefield acceleration, for instance, a laser pulse drives a large-amplitude plasma wave that can trap and accelerate electrons to high energies. When driven by conventional pulses, laser wakefield accelerators face a limitation referred to as ``dephasing,'' where the high-energy electrons outrun the accelerating phase of the plasma wave and begin to decelerate. An arbitrary-velocity pulse with $v_\mathrm{a} = c$ can drive a plasma wave with a phase velocity equal to $c$, which prevents electrons from outrunning the accelerating phase. The original concept, based on flying-focus pulses \cite{Caizergues2020,palastro2020}, used an axiparabola \cite{smartsev2019} to create an extended focal range in a uniform plasma. This introduces two challenges: (i) the spot size of the pulse evolves along the focal range and (ii) a large volume of plasma is needed to mitigate refraction. An arbitrary-velocity pulse in a plasma waveguide could overcome both of these challenges: the spot size would be relatively fixed at $w$ and the waveguide would require a much smaller volume of plasma. In addition, a train of arbitrary-velocity pulses with $v_\mathrm{a} = c$ and a recurrence period $T_\mathrm{R} = 2\pi/\omega_\mathrm{p0}$ (or integer multiples thereof) could mitigate dephasing in multi-pulse laser wakefield acceleration \cite{Hooker2014}.

The analysis and simulations presented here considered longitudinally uniform plasma waveguides. Arbitrary-velocity pulses may also be created in a longitudinally structured plasma waveguide or even remove the need for such a structure. A corrugated waveguide, in particular, would provide additional control by allowing for luminal or subluminal phase velocities $v_\phi \leq c$ \cite{Layer2007}. A guided laser pulse with both $v_\phi = c$ and $v_\mathrm{a} = c$ would overcome the ``pulse length dephasing'' limitation on the energy gain in direct laser acceleration of electrons \cite{Palastro2008}. 
One approach to high-power THz generation in a plasma relies on matching the velocity of the source---a ponderomotively driven current---to the phase velocity of the THz radiation \cite{Antonsen2007}. This is achieved in a corrugated waveguide by matching the subluminal group velocity of a conventional laser pulse to the subluminal phase velocity of a THz mode in the waveguide. An arbitrary-velocity intensity peak could instead drive a superluminal current with a velocity matched to the usual superluminal phase velocity of THz, obviating the corrugated structure. 

The spatiotemporal profile of the moving intensity peaks could be further structured by extending the analysis and simulations beyond equal-amplitude modes with $\ell = 0$. As with a Fourier series, using modes with different amplitudes would allow for temporal shaping of the intensity peaks. Finally, superposing modes with frequencies that also depend on $\ell$ would enable guided light springs, arbitrary-velocity light springs, or more intricate motion of the peak intensity in both $\theta$ and $z$. 

\begin{acknowledgments}
The authors would like to thank M. Piccardo, D.H. Froula, D. Ramsey, K. Weichman, and A. Konzel for insightful discussions. The work of J.P.P., K.G.M., A.L.E., and L.S.M. is supported by the Office of Fusion Energy Sciences under Award Numbers DE-SC0021057, the Department of Energy National Nuclear Security Administration under Award Number DE-NA0004144, the University of Rochester, and the New York State Energy Research and Development Authority. This report was prepared as an account of work sponsored by an agency of the US Government. Neither the US Government nor any agency thereof, nor any of their employees, makes any warranty, express or implied, or assumes any legal liability or responsibility for the accuracy, completeness, or usefulness of any information, apparatus, product, or process disclosed, or represents that its use would not infringe privately owned rights. Reference herein to any specific commercial product, process, or service by trade name, trademark, manufacturer, or otherwise does not necessarily constitute or imply its endorsement, recommendation, or favoring by the US Government or any agency thereof. The views and opinions of authors expressed herein do not necessarily state or reflect those of the US Government or any agency thereof.
\end{acknowledgments}

\appendix
\section{Paraxial Intersection Wavelengths}
To recast Eq. \eqref{eq:parax} in a form that is more amenable to experimental design, it is convenient to parameterize the profile of a parabolic plasma channel as follows: 
\begin{equation}\label{eq:altden}
n(r) = n_0+\left(\frac{r}{R_\mathrm{d}}\right)^2n_0,
\end{equation}
where $n_0$ is the on-axis electron density and $R_\mathrm{d}$ is the radius at which $n = 2n_0$. In terms of the matched spot size $w$, $R_\mathrm{d} = (\pi r_\mathrm{e} n_0)^{1/2} w^2$, where $r_\mathrm{e}$ is the classical electron radius. The required vacuum wavelength for each mode is then 
\begin{equation}\label{eq:paraxEX}
\lambda_{q\ell} \approx \lambda_0 - \frac{(2q + |\ell|)}{2\pi^{3/2}} \frac{v_\mathrm{a}}{(v_\mathrm{a} - v_{\mathrm{g}0})} \frac{(r_\mathrm{e} n_0)^{1/2}\lambda_0^3}{R_\mathrm{d}},
\end{equation}
where $\lambda_0 = 2\pi c/\omega_0$.

\section{Step-Index Plasma Channel}
While laser-heated plasmas typically form parabolic plasma channels, some applications benefit from channels with hollow or step function profiles \cite{Chiou1995,Schroeder1999,Schroeder2016}. The profile of a step-index channel can be parameterized as follows:
\begin{equation}\label{eq:step}
\omega_\mathrm{p}^2(r) =  
\bigg\{
\begin{array}{ll} 
      \omega_{\mathrm{p}0}^2 & r\leq R \vspace{4pt} \\ 
      \omega_{\mathrm{p}1}^2  & r>R \\
\end{array}, 
\end{equation}
where $R$ is the channel radius and for a hollow plasma channel $\omega_\mathrm{p0}^2 = 0$. The corresponding $A_{q\ell}(r)$ are given by
\begin{equation}\label{eq:Astep}
A_{q\ell}(r) = \Bigg\{
\begin{array}{ll} 
      J_\ell(r/w_{q\ell}) & r\leq R \vspace{4pt} \\ 
      \frac{J_\ell(R/w_{q\ell})}{K_\ell(R/u_{q\ell})}  K_\ell(r/u_{q\ell})   & r>R \\ 
\end{array},       
\end{equation}
where the $J_\ell$ are Bessel functions of the first, the $K_\ell$ are modified Bessel functions of the second kind, and $u_{q\ell}^{-2} \equiv w_{q\ell}^{-2} - (\omega_\mathrm{p1}^2 - \omega_\mathrm{p0}^2)/c^2$. The $w_{q\ell}^{-2}$ are determined by all possible solutions to 
\begin{equation}\label{eq:wstep}
 \frac{J_{\ell+1}(R/w_{q\ell})}{J_{\ell}(R/w_{q\ell})} = \frac{w_{q\ell}}{u_{q\ell}} \frac{K_{\ell+1}(R/u_{q\ell})}{K_{\ell}(R/u_{q\ell})}.
\end{equation}
When $\Delta \equiv (\omega_\mathrm{p1}^2 - \omega_\mathrm{p0}^2)^{1/2}(R/c) \gg 1 $, the $w_{q\ell}^{-2}$ can be approximated as
\begin{equation}\label{eq:wstep}
\frac{1}{w_{q\ell}^2} \approx \frac{\pi^2}{4R^2}\left[\left(\frac{\Delta + |\ell| - \tfrac{3}{2}}{\Delta + |\ell| + \tfrac{1}{2}}\right)(2q+|\ell|+\tfrac{3}{2})^2 - \frac{(4\ell^2 -1)}{\pi^2} \right].
\end{equation}
In contrast to the parabolic plasma channel [Eq. \eqref{eq:wpara}], the $w_{q\ell}^{-2}$ for a step-index channel depend nonlinearly on $q$ and $\ell$. 

An arbitrary-velocity intensity peak in a step-index plasma channel will only exhibit recurrences when it is composed of modes with $\ell = 0$. This is because the relative frequencies required for the moving intensity peak can only be commensurate if the $4\ell^2/\pi^2$ term in Eq. \eqref{eq:wstep} vanishes. When $\ell = 0$ and the paraxial approximation is valid,
\begin{equation}\label{eq:steprat}
\frac{\Omega_{q0}}{\Omega_{p0}}  \approx \frac{2q^2 + 3q}{2p^2 + 3p},
\end{equation}
which is indeed rational. As a result, the moving intensity peak exhibits recurrences with a period $T_\mathrm{R} = 2\pi (2q^2 + 3q)/\Omega_{q0}$ or 
\begin{equation}\label{eq:Trecurstep}
T_\mathrm{R} \approx \frac{8 \omega_0 R^2}{\pi c^2} \bigg| \frac{v_\mathrm{a} - v_\mathrm{g0}}{v_\mathrm{a}} \bigg|,
\end{equation}
where $\Delta \gg 1$ has been assumed. Similary, the effective duration of the intensity peak scales roughly as
\begin{equation}\label{eq:Tdurstep}
\tau \sim \frac{4 \omega_0 R^2}{\pi c^2(2q_\mathrm{max}^2 + 3q_\mathrm{max})} \bigg| \frac{v_\mathrm{a} - v_\mathrm{g0}}{v_\mathrm{a}} \bigg|.
\end{equation}
While the step-index channel does allow for recurrences, the nonharmonicity of the relative frequencies results in a primary intensity peak surrounded by less-structured, lower-amplitude peaks. Thus, the intensity contrast in a step-index channel is worse than in a parabolic plasma channel.

\section{Quartic Perturbation}
This appendix examines how the modal dispersion relation and intensity recurrences are modified when a predominantly parabolic plasma channel is weakly quartic. With the inclusion of a quartic perturbation, the profile of an otherwise parabolic channel can be parameterized as follows:
\begin{equation}\label{eq:quart}
\omega_\mathrm{p}^2(r) = \omega_{\mathrm{p}0}^2 + \frac{4c^2}{w^4}r^2 + \varpi\frac{c^2}{6w^6}r^4, 
\end{equation}
where the dimensionless parameter $\varpi \ll 1$ quantifies the strength of the perturbation. The lowest-order correction to the $w_{q\ell}^{-2}$, denoted by $\delta w_{q\ell}^{-2}$, is calculated using first-order perturbation theory:
\begin{equation}\label{eq:pert1}
\delta w_{q\ell}^{-2} = 
\frac{\varpi}{6w^6} \frac{\int r^4 A_{q\ell}^2(r) d^2r}{\int A_{q\ell}^2(r) d^2r},
\end{equation}
where the $A_{q\ell}(r)$ are given by Eq. \eqref{eq:Apara}. Upon performing the integral, one finds
\begin{equation}\label{eq:pert2}
\delta w_{q\ell}^{-2} = \frac{\varpi}{4w^2}\left[ q^2 + (|\ell| + 1)q + \frac{1}{6}(|\ell| + 1)(|\ell| + 2) \right].
\end{equation}
Substituting $w_{q\ell}^{-2} \rightarrow w_{q\ell}^{-2} + \delta w_{q\ell}^{-2}$ in Eq. \eqref{eq:dispersion} corrects the modal dispersion relation to first order in the quartic perturbation.

The presence of a quartic term in the channel profile makes the relative frequencies slightly anharmonic, even if the paraxial approximation is valid. For consistency with Section IV, consider the case of $\ell = 0$. To first order in the quartic perturbation $W_{q\ell}^{-2} = (2c/\omega_0w)^2[(2+\varpi/16)q + (\varpi/16)q^2]$, such that
\begin{equation}\label{eq:pertrat}
\frac{\Omega_{q0}}{\Omega_{10}}  \approx q + \frac{\varpi}{32}(q^2-q).
\end{equation}
The perturbation term in Eq. \eqref{eq:pertrat} (i.e., the term ${\propto} \varpi$) makes the ratio non-integer and the frequencies anharmonic. Over a long enough time, the perturbation will cause the recurrences to phase mix away. This occurs when two modes are $\pi$ out phase after a number of recurrence periods---that is, when $\mathrm{mod}[(\Omega_{q0} - \Omega_{10})N_qT_\mathrm{R},2\pi] = \pi$, where $N_q$ is the number of recurrence periods. Substituting in Eq. \eqref{eq:pertrat} yields
\begin{equation}\label{eq:Nq}
N_q = \frac{16}{\varpi(q^2-q)}.
\end{equation}
As an example, with $q = 5$ and $\varpi = 1/5$, the recurrences would persist for $N_5 = 4$ periods. 

\section{Simulation Details}
The simulations presented in this work were performed using \textsc{qpad} \cite{Li2021,Li2022}. \textsc{qpad} is a quasi-static particle-in-cell code that takes advantage of the large separation in time scales between a laser period $2\pi/\omega_0$ and plasma period $2\pi/\omega_{\mathrm{p}}$ or laser pulse duration $T$. The large separation allows for the use of cycle-averaged equations of motion, where the plasma electrons evolve in response to the ponderomotive force of the laser pulse and the electrostatic fields that it drives. In addition, the evolution of the laser pulse can be reduced to the evolution of its envelope, without the need to resolve the laser period. This provides a large computational savings ${\sim}\mathcal{O}(\omega_0^3/\omega_\mathrm{p}^3,c^2\omega_0^2/\omega_\mathrm{p} w^2)$ compared to traditional particle-in-cell methods.

\textsc{qpad} uses a moving frame and is discretized in the quasi-3D geometry, which decomposes the fields of the laser pulse and plasma into a truncated expansion of azimuthal modes. In the simulations presented here, the moving-frame window was 10.3~ps $\times$ $96 \; \mu$m (12288 $\times$ 8192 cells) in $\xi$ and $r$, respectively, and only the zeroth-order azimuthal mode was used. The step in $z$ ($s$ in Ref.~[\onlinecite{Li2021}]) was $20 \; \mu$m. The electrons were represented by 32 particles per cell. The longitudinal profile of the plasma had an initial $80$-$\mu$m upramp but was otherwise uniform. The temporal profile of the laser pulse consisted of a 1.8-ps rise, a 6.4-ps plateau, and a 1.8-ps fall, yielding an intensity FWHM of 7.7~ps. The rise and fall were 6th-order polynomials whose derivatives vanish at their end points.

\bibliography{main} 

\end{document}